\documentstyle[12pt]{article}
\def\be{\begin{equation}}
\def\ee{\end{equation}}
\def\bea{\begin{eqnarray}}
\def\eea{\end{eqnarray}}

\author{Hans - J\"urgen Schmidt}

\title{Inhomogeneous Cosmological Models with Flat Slices Generated
from the {\sc Einstein--de  Sitter} Universe }

\date{}
\begin{document}
\maketitle

\centerline{Universit\"at Potsdam, Institut f\"ur Mathematik, Am
Neuen Palais 10} 
 \centerline{D-14469~Potsdam, Germany,  E-mail:
 hjschmi@rz.uni-potsdam.de}

\begin{abstract}
A family of cosmological models
 is considered which in a certain synchronized system of
reference possess flat slices
$t=$  const. They are generated from the EINSTEIN-DE SITTER universe 
by a suitable
transformation. Under physically
reasonable presumptions
 these transformed models fulfil certain energy conditions. 

\bigskip

Es wird eine
Familie kosmologischer Modelle betrachtet, die in einem gewissen 
synchronisierten
Bezugssystem flache
Schichten $t=$  const. besitzen. 
Sie werden mittels einer geeigneten Transformation aus dem
EINSTEIN-DE SITTER-Universum erzeugt. 
Diese transformierten Modelle erf\"ullen unter
physikalisch sinnvollen Voraussetzungen gewisse Energiebedingungen.
\end{abstract}

\section{Introduction}%1

In WAINWRIGHT (1981),  SCHMIDT (1982)  and references cited 
there a  class of
inhomogeneous cosmological models
 is considered which have the following property: there
exists a synchronized system of reference of such a 
kind that the slices $t=$ const. are
homogeneous manifolds. Here
 we consider a special family of such models which possess flat
 slices. 
To this end we use the transformation formalism developed in
 SCHMIDT (1982).
Additionally, we require that these transformations
 leave two coordinates unchanged; this
implies the existence of a 2-dimensional Abelian group of motions. 
(A similar requirement is
posed in WAINWRIGHT (1981), too.)  Starting from a FRIEDMAN 
universe, 
 we investigate
whether the energy inequalities are fulfilled in the transformed model, too. 
In general this fails
to be the case, 
but starting from the EINSTEIN - DE SITTER universe 
(EINSTEIN 1932; cf. also
TOLMAN 1934) and
 requiring perfect fluid for the transformed model, the energy inequalities
in the initial model imply their validity in the transformed model. These
 statements answer
partially a question posed by TREDER.

\section{Models with flat slices}

The transformation formalism of SCHMIDT (1982) restricted to BIANCHI 
type I 
reads as
follows: using the same notations, we have $A_i^a = \delta_i^a$,
$\omega^a = dx^a$ and
\be%1
     ds^2 = - dt^2 + g_{ab}(t) dx^a dx^b       % (1)
\ee
as the initial hypersurface-homogeneous model. 
Now let us consider the time-dependent
transformation $x^a_t (x^i, t)$, where for 
each $t$ it has to be a diffeomorphism of $R^3$. 
Then one obtains
\be%2
 g_{0\alpha}    =  - \delta_{0 \alpha} , \qquad g_{ij} = g_{ab}(t)
 x^a_{t,i} x^b_{t,j}  % (2)
\ee
as the transformed model. (It  is no restriction to insert 
$g_{ab}(t) = \delta_{ab}$  in (1), i.e. to start from
MINKOWSKI space.)

In the following we consider
 only transformations which leave two  coordinates unchanged, i.e.
(now writing $t$, $x$, $y$, $z$ instead of $x^0, \dots x^3$  resp.)
 transformations which read as follows 
\be%3
     x_t(x, t) \,  , \qquad   y_t = y \,  , \qquad z_t = z \, .   %(3)
\ee
(These we shall
 call $x$-transformations.) 
Using $x$-transformations,
 the KILLING vectors   $\partial/\partial y$  
and     $\partial/\partial z$    of the initial model remain KILLING 
vectors. 
They form a 2-dimensional Abelian
group of motions. All others 
(including the rotation $ z(\partial/\partial y) - y (  \partial/\partial
z) $)   
may fail  to remain
KILLING vectors. On the contrary to the general case, the 
$x$-transformed models depend
genuinely on the initial ones. In the following the 3-flat FRIEDMAN 
universe shall  be used  as
initial model, i.e.
\be%4
 g_{ab} (t) = \delta_{ab} K^2(t) \qquad {\rm with} \qquad K(t) = t^\tau \, .
\ee
Together with (1) one obtains
\be%5
\kappa \mu = \kappa T_{00} = 3\tau^2 / t^2 \, , \quad
p=T^2_2 = \alpha \mu 
    \quad {\rm with} \quad
 \alpha = \frac{2}{3\tau} -1 \, .
\ee
Now, inserting (4) in (1) and transforming 
to (2) with restriction (3) one obtains for the metric
of the $x$-transformed
model
\be%6
     g_{11} = t^{2\tau} \cdot 
    (x_{t,1})^2 \equiv   t^{2\tau} \cdot  h(x, t) \,  , \quad 
 g_{22} = g_{33} =     t^{2\tau} \, , \quad 
 g_{\alpha \beta} = \eta_{\alpha \beta}
 \   {\rm    else} \, . 
\ee
This metric belongs
 to the so-called
 SZEKERES class,  cf.  SZEKERES (1975).  Defining $a(x, t)$
by
\be%7
     g_{11} =   e^{2     a(x, t) } \,   ,
\ee
a coordinate  transformation $\tilde x(x)$ yields
 $ a(x, 1) = 0$. If $v = a_0 \, t$,  then we have
\be%8
     a(x, t) = \int_1^t    v(x, \tilde t) \cdot \tilde t^{-1} d \tilde t \,
,    
\ee
$v(x, t)$
 being an arbitrary  twice continuously differentiable function
 which may be singular at $t
= 0$ and $t = \infty$.
 (The initial model is included by setting $x_t = x$, hence 
$a = \tau \ln t$, $v = \tau$.)

For
$\tau \ne 0 $
 different functions $v$ correspond to the same model  only if
 they are connected by a
translation into $x$-direction. 
Inserting  (6) with (7) and (8) into the EINSTEIN equations one
obtains the energy-momentum tensor
\bea %9
\kappa T_{00} = ( \tau^2 + 2 v \tau   )/t^2 \, ,
    \nonumber \\
\kappa T^1_1 =  (2  \tau - 3 \tau^2   )/t^2 \, ,
    \nonumber \\
\kappa T^2_2 = \kappa T^3_3 = -v_{,0} t^{-1} +
 (  \tau -  \tau^2   - v\tau - v^2 + v  )/t^2 \, ,
    \nonumber \\
\kappa T_{\alpha \beta} = 0
 \qquad {\rm else   \qquad  and}     \nonumber \\
\kappa T = \kappa T^\alpha_\alpha = - 2 v_{,0} t^{-1} -
 2 ( 3 \tau^2 - 2 \tau +2 v\tau + v^2 - v  )/t^2 \, .
\eea
The question, in which cases (6) represents a usual
  hypersurface-homogeneous model, can 
be answered as follows: metric (6) is a FRIEDMAN universe, 
if and only if $h_{,0} = 0$. 
For this case
it is isometric to the initial model. Metric (6) is a 
hypersurface-homogeneous model, if
 and only
if  functions $A$ and $B$
 exist for which holds $h(x, t) =  A(x) \cdot  B(t)$.
 Because of $h > 0$  this is
equivalent to   $ h_{,01} \cdot  h = h_{,0} \cdot h_{,1}$.
 In this case it is a BIANCHI type I  model.

\section{Energy inequalities}%3

In this section it shall be discussed, 
in which manner the geometrically defined models
described by equs. (6), \dots ,  (9)
 fulfil some energy conditions. Here we impose the following
conditions: each observer measures non-negative energy density, 
time- or light-like energy flow
and space-like tensions which are not greater 
than the energy density. In our 
coordinate   system
these conditions are expressed by the following inequalities
\be%10
     T_{00} \ge \vert T^1_1 \vert \, ,  
\ee
\be%11   
     T_{00} \ge \vert T^2_2 \vert \, ,  
\ee
\be%12  
 {\rm   and } \qquad   T \le 0 \, . 
\ee
For the initial model this means $\tau =   0$  or $ \tau \ge 1/2$, i.e.
MINKOWSKI 
space or $-1 < \alpha \le 1/3$. Using the energy-momentum tensor 
(9), equ. (11) and (12) read
\be%13
 v_{,0} \, t \ge -v^2 - (3\tau - 1)  v + \tau - 2 \tau^2       \, ,
\ee
\be%14
 v_{,0} \, t \le -v^2 + (\tau +1) v + \tau   \qquad   {\rm   and }
\ee
\be%15
 v_{,0} \, t \ge -v^2  - (2 \tau -1) v + 2\tau - 3 \tau^2     \, .
\ee

Now, if $\tau  < 0$, i.e. $\alpha  < -1$, then (10)  
reads $v \le \tau -1$;  together with (14) one obtains 
$v_{,0} \, t \le \tau - 2$. This
 implies the existence of  a $\tilde t > 0$ 
with   $v(\tilde t) \ge -1$ in contradiction to (10).
 Therefore, an initial model with $\tau  < 0 $ (which itself
 contradicts the energy inequalities) cannot produce 
transformed  models which always fulfil them.

If $\tau = 0$, then (13) and (14) imply  $v_{,0} \, t  = v - v^2$. 
This   equation has the solutions $v = 0$  and
$ v = t(t + C)^{-1} $  with 
arbitrary $C(x) \ge 0$. This yields $a = 0$ and $a = \ln (t + C) - 
\ln (1+  C) $  with (7)
and $g_{11} = 1$ and 
$ g_{11} = (t + C)^2 \cdot  (1 + C)^{-2}$   resp. with (8). Then 
(6) shows that this is the 
MINKOWSKI space itself. Therefore, the MINKOWSKI space does
 not produce  any new  models.

Finally, if $\tau > 0$, hence $\alpha  > -1$, (10) then reads 
 $v \ge {\rm  max} (\tau -   1, 1 - 2\tau)$. 
A lengthy   calculation
shows which transformed models fulfil the energy inequalities. 
For each $\tau > 0$ models exist
 which do and models which  do not fulfil them.

\section{Perfect fluid models}%4

The situation described above changes if
 one requires that the transformed model consists of
 perfect fluid with an equation of state. The velocity vector must be 
 $(1, 0, 0, 0)$  and
\be%16
\kappa(T^1_1 - T^2_2 ) \equiv 
\frac{
h_{,0}
}{
(1+\alpha) t \cdot h
}
+ \frac{1}{2 \sqrt h}
\left(
\frac{  h_{,0}  }{\sqrt h}
\right)_{,0}
=0
\ee
must be fulfilled. If $ f = h_{,0}h^{-1/2}$, then (16) reads
$$
\frac{f}{(1+\alpha) t}
+
\frac{f_{,0}}{2}
= 0 \, ,
$$
hence $f_{,0} f^{-1}$
  does not depend on $x$,
 therefore $f_{,01}\cdot f = f_{,0} \cdot  f _{,1}$, 
and we can use 
the ansatz   $f =   a(t) \cdot b(x)$.
Inserting this    into  (16), one obtains
\be%17
h= \left[ b(x) \cdot 
 t^{(\alpha -1)(\alpha +1)}
 + c(x) \right]^2
\, , \quad {\rm    where } \quad 
  c(x) = \sqrt{ h(x,0)}
\ee
with arbitrary non-negative functions $b$ and $c$
 fulfilling $b(x) + c(x) > 0$ for all $x$. For energy
density and pressure we then obtain
\bea%18
\kappa \mu = \frac{4}{3(1+\alpha)^2 t^2}
 -
\frac{4(1-\alpha) \cdot b
}{3(1+\alpha)^2 t (bt + ct^{2/(1+\alpha)}) }  \nonumber \\
\kappa p = \frac{4\alpha}{3(1+\alpha)^2 t^2} \, .
\eea
An equation of state means that $p$ uniquely depends on $\mu$.
 This takes  place, if and only if $\alpha =  0$ or
 $b/c =$  const. The latter is equivalent to the
 hypersurface-homogeneity of the model and is of lower interest here.
For $\alpha =   0$      the initial
model is the dust-filled EINSTEIN-DE SITTER model 
(EINSTEIN 1932). With (18) we obtain
 $p =  0$  and $\mu \ge   0$  and may
formulate:   

\noindent 
If the EINSTEIN-DE SITTER
 universe is $x$-transformed into a perfect fluid model, then
 this model also contains dust and fulfils the energy conditions.

These models have the following form: 
inserting (17) with $\alpha =  0$  into (6) we get a dust-filled model
\be%19
ds^2 \,  = \,  -dt^2 \,  + \,  t^{4/3} \, 
\{ \, 
 [ \, b(x)/t + c(x) \, ]^2 \, dx^2 \,  + \,  dy^2 \,  + \,  dz^2 \,   \} 
\ee
with arbitrary $b$, $c$ as before.
Equ. (19) is contained in SZEKERES (1975) as case (iii), but the 
parameter $\epsilon$ used there may be
non-constant.

\noindent 
As an illustration we give two examples of  this model (19):

\noindent 
1.   If $b =1$, $c > 0$, then
$$
\kappa \mu = \frac{4c}{3t(1+ct)} \, ,
$$
hence the density contrast at two different values $x_1$, $x_2$
 reads
$$
 \frac{\mu_1}{\mu_2} =
  \frac{c_2}{c_1}\cdot  \frac{1 + c_1 t}{1 + c_2 t} 
$$
and tends to 1 as $t \to \infty$. This shows that one 
needs additional presumptions if one wants to
prove an amplification of initial density fluctuations.

\noindent
2. If $b = 1$, $c = 0$, then (19) is the KASNER 
vacuum solution. If now $c$ differs from zero in the
neighbourhoods of two values     $x_1$, $x_2$,  then 
the model is built up from two thin dust slices and
KASNER-like vacuum outside them. 
The invariant distance of  the slices is
$$
A t^{-1/3} + B t^{2/3}
$$
with certain positive constants $A$  and $B$.  
This looks like gravitational repulsion, because the
distance has a minimum 
at a positive value. But the $t^{-1/3}$-term is due to
 the partizipation of the
slices in cosmological expansion and 
the remaining $t^{2/3}$-term is due to an attractive
gravitational force in parabolic  motion.

\section{Conclusion}%5

The transformation of a hypersurface-homogeneous cosmological
 model
 considered here firstly preserves all inner properties (expressed by 
the first 
fundamental form) of the slices $t = $ const., 
and secondly the property that $t$  is a synchronized
time, but may change all other outer
 properties (essentially expressed by the second fundamental
form). The investigations of sections 3. and 4. show that energy 
conditions are preserved 
under  very special presumptions only.

One may consider these transformations as a guide in the search
 for new exact solutions of
EINSTEIN's  field equations. The new models  are
 {\it close}  to the initial hypersurface-homogeneous
ones, if the transformation is close
 to the identical one. Thus, one can perturb a BIANCHI model
 with exact solutions without use of  any  (uncertain) approximations.

\noindent 
I want to thank 
Professor TREDER and Dr. v. BORZESZKOWSKI for helpful 
discussions.

\section*{References}

\noindent
EINSTEIN, A., W. DE SITTER: 1932, Proc. Nat. Acad. Sci. {\bf 18}, 213.

\noindent
SCHMIDT, H.-J.: 1982, Astron. Nachr. {\bf  303}, 231.\footnote{This
 is a misprint in the original, the cited paper is at page 227
 and not at page 231; I thank Malcolm MacCallum for his note.} 

\noindent
SZEKERES, P.: 1975, Commun. Math. Phys. {\bf 41}, 55.

\noindent
TOLMAN, R. C.: 1934, Relativity Thermodynamics and Cosmology, \S 164, Oxford.

\noindent
WAINWRIGHT, J.: 1981, J. Phys. A {\bf 14}, 1131.

\bigskip

\noindent
(Received 1982 January 22)

\medskip

\noindent 
{\small {In this reprint (done with the 
kind permission of the copyright owner) 
we removed only obvious misprints of the original, which
was published in Astronomische Nachrichten:   
 Astron. Nachr. {\bf 303} (1982) Nr. 5, pages 283 - 285;  
  Author's address that time:  
Zentralinstitut f\"ur  Astrophysik der AdW der DDR, 
1502 Potsdam--Babelsberg, R.-Luxemburg-Str. 17a.}}

\end{document}